\documentclass[aps,pra,superscriptaddress,twocolumn]{revtex4-2}
\usepackage{amssymb}
\usepackage{mathptmx}
\usepackage{amsmath}
\usepackage{graphicx}% Include figure files
\usepackage{dcolumn}% Align table columns on decimal point
\usepackage{bm}% bold math
\usepackage[colorlinks=true,allcolors=blue]{hyperref}% add hypertext capabilities
\usepackage[mathlines]{lineno}% Enable numbering of text and display math
\usepackage[dvipsnames]{xcolor}
\usepackage{xcolor}
\DeclareMathAlphabet{\mathcal}{OMS}{cmsy}{m}{n}

\newcommand{\ket}[1]{|#1\rangle}
\newcommand{\bra}[1]{\langle#1|}
\newcommand{\mean}[1]{\langle#1\rangle}
\newcommand{\hatd}[1]{\hat{#1}^\dag}
\newcommand{\proj}[1]{|#1\rangle\!\langle#1|}
\newcommand{\tran}[2]{|#1\rangle\!\langle#2|}
\newcommand{\half}[0]{\tfrac{1}{2}}
\newcommand{\Nbos}[0]{\bar{N}}

\newcommand{\tr}[1]{\mathrm{Tr}#1}

\definecolor{myblue}{RGB}{0, 122, 255}

\begin{document}
\title{High fidelity macroscopic superposition states via shortcut to adiabaticity}

\author{Mehdi Aslani}
\email{mehdiaslanimk@gmail.com}
\affiliation{Department of Physics, University of Isfahan, Hezar-Jerib, Isfahan 81746-73441, Iran}
\affiliation{Quantum Optics Group, Department of Physics, University of Isfahan, Hezar-Jerib, Isfahan 81746-73441, Iran}
 
\author{Vahid Salari}%
\affiliation{Department of Physics and Astronomy, University of Calgary, Calgary T2N 1N4, Alberta, Canada}
\affiliation{Institute for Quantum Science and Technology (IQST), University of Calgary, Calgary T2N 1N4, Alberta, Canada}
  
\author{Mehdi Abdi}%
\email[]{mehabdi@gmail.com}
\affiliation{Wilczek Quantum Center, School of Physics and Astronomy,
Shanghai Jiao Tong University, Shanghai 200240, China}
%\affiliation{Department of Physics, Isfahan University of Technology, Isfahan 84156-83111, Iran}

\date{\today}

\begin{abstract}
A shortcut to adiabatic scheme is proposed for preparing a massive object in a macroscopic spatial superposition state.
In this scheme we propose to employ counterdiabatic driving to maintain the system in the groundstate of its instantaneous Hamiltonian while the trap potential is tuned from a parabola to a double well.
This, in turn, is performed by properly ramping a control parameter.
We show that a few counterdiabatic drives are enough for most practical cases.
A hybrid electromechanical setup in superconducting circuits is proposed for the implementation.
The efficiency of our scheme is benchmarked by numerically solving the system dynamics in the presence of noises and imperfections.
The results show that a mechanical resonator with very high fidelity spatially distinguishable cat states can be prepared with our protocol.
Furthermore, the protocol is robust against noises and imperfections.
We also discuss a method for verifying the final state via spectroscopy of a coupled circuit electrodynamical cavity mode.
Our work can serve as the ground stone to feasibly realize and verify macroscopic superposition states in future experiments.
\end{abstract}

\maketitle
%\tableofcontents

%§§§§§§§§§§§§§§§§§§§§§§§§§§§§§§§§§§§§
\section{Introduction}
\label{sec:intro}
In recent decades two aspects of the quantum mechanics has become growingly prominent:
First, its application in the technology and advantages that it brings over the classical competitors, e.g. in enhanced sensing schemes and secure communications~\cite{Gisin2007, Degen2017, Pirandola2018, Georgescu2014, Ladd2010}.
And in the foundation of quantum theory itself, where several questions still need to be addressed.
Crucially, those questions about where the quantum realm meets the classical mechanics and nonclassicality of dynamics~\cite{Modi2012, Marletto2020, Smirne2018, Abdi2023}.
Among the other aspects, it still remains unclear whether it is the system size, its number of degrees of freedom, or its mass that determines the limit were one \textit{must} invoke the quantum theory for understanding its dynamics~\cite{Castagnino2005, Kiesel2012, Nimmrichter2013, Gittsovich2015}.
One of the well-established approaches for addressing this issue are the theoretical extensions that predict unconventional mechanisms for decoherence~\cite{Bassi2013}.
Such theories usually provide a decoherence rate related to the size, mass, or the degrees of freedom of the system.
Suggesting that the quantum states in a larger system loses its coherence faster~\cite{Arndt2014}.

Performing experiments nonetheless are necessary for testing the validity of these theories.
It usually requires the ability of preparing a massive object in a superposition state~\cite{Schlosshauer2005, Romero2011, Abdi2016, Hu2022} or equivalently matter wave interferometry with large objects~\cite{Brezger2002, Nimmrichter2011, Cernotik2019, Xuereb2015}.
Given the sensitivity of such systems it is necessary to be able to prepare such states with very high fidelity.
Nonetheless, massive objects accessible in current experimental opto- and electro-mechanical systems are also subject to a tremendous amount of thermal noise.
Therefore, one must conceive proper approaches where the nonclassical states can be achieved with high fidelity, and thus, making possibility for the observation of unconventional decoherence effects.
Massive objects in quantum superposition could also prove useful for enhanced sensitivity in force measurement~\cite{Yan2023, Li2021, Motazedifard2016, Hempston2017, Asjad2023, Ranjit2016, Bemani2022}.
Therefore, various proposals have been put forth for preparing macroscopic objects in spatially distinguishable superposition states; by dissipative state preparation~\cite{Abdi2016, Tan2013}, hybrid system manipulation~\cite{Abdi2015a, Asadian2014, Liao2016, Abdi2017, Wang2023}, measurement induced~\cite{Hoff2016, Clarke2018, Sekatski2014}, and adiabatic processing~\cite{Buchmann2012, Chen2021, Qi2022}.
%Either of these techniques have their own advantages and disadvantages.
%Even though engineering the environment could provide a robust and long-lived state the fidelity of resulting states are not usually satisfactory.
%On the other hand, the transient states obtained from other approaches usually give little opportunity for the validation.

Here, we investigate a scheme in which shortcut to adiabaticity is employed for the rapid and high fidelity preparation of a massive macroscopic object in a superposition state~\cite{GueryOdelin2019}.
The cat state is realized by preparing the massive system in the ground state of a double-well (DW) potential.
In the scheme we propose, a mechanical mode is cooled down to its ground state while oscillating in an almost harmonic trap with a weak Duffing nonlinearity.
Then the potential is twisted into a DW by applying an external anti-parabola potential.
By retaining the system in the ground state of the instantaneous potential during the process, the mechanical resonator in the desired quantum state can be achieved.
However, this is challenged by two effects:
On the one hand, the thermal noise excites the system to other states making a mixed incoherent state.
Such thermalization effects become growingly prominent as the lowest energy gap grows smaller with the formation of the DW.
On the other hand, speeding up the process by employing faster ramps results in the diabatic transitions in the system through Landau-Zener effect which again prohibit formation of the desirable superposition state.
To overcome this, we propose to accelerate the procedure by employing counter-diabatic drives~\cite{Chen2010, Garaot2013, Campo2013}.

We use an approximate version of \textit{transitionless quantum driving}~\cite{Berry2009, Hatomura2018}, where only a few substantial diabatic transitions are compensated for through appropriately driving modes of a coupled cavity.
Therefore, both the energy costs and the experimental feasibility are significantly relaxed.
We benchmark our protocol by computing the final state fidelity through our numerical solutions to the quantum optical master equation considering realistic noise effects.
By optimizing the required resources, we show that the groundstate of the DW potential is attainable with a high fidelity by only employing a limited number of counter-drive fields, or equivalently limited number of cavity modes.
We compare our results with the states obtained via a simple adiabatic passage protocol in the same conditions and show that the protocol performance is significantly better.
Then different protocol scenarios as well as imperfections are studied.
The latter includes the asymmetry in the potential that breaks the parity symmetry of the states as well as a finite thermal occupation as the starting point of the protocol.
Eventually, we propose a readout technique through spectroscopy of a coupled cavity mode for verifying the state of the mechanical resonator.

The paper is organized as follows:
In the next section we discuss the preliminary theoretical aspects of our work, including the model, the protocol, and the proposed setup.
In Sec.~\ref{sec:results} the numerical results are presented for an adiabatic process and the protocol with shortcut to adiabaticity.
Sec.~\ref{sec:tomography} puts forward a method for verifying the prepared state.
The paper is concluded by Sec.~\ref{sec:conclusion}.

%§§§§§§§§§§§§§§§§§§§§§§§§§§§§§§§§§§§§
\section{Theory}
\label{sec:theory}
When a control parameter of a quantum system changes over time it can modify the Hamiltonian and consequently its corresponding eigenstates.
If the change is performed slowly enough, a system prepared in one of its eigenstates, e.g. groundstate, retains that status without occupying other eigenstates.
This indeed is the so called quantum adiabatic theorem and it is commonly used in quantum information processing~\cite{Kato1950, Albash2018, Rezaei2022, Venuti2017, Hegade2021, Xiang2023}.
Although one in principle should perfectly achieve the desired state by changing a Hamiltonian through arbitrarily long processes, the environmental noises and dissipations are prohibitive.
Therefore, it is necessary to design processes fast enough that the decohering effects are minimal, yet the adiabatic nature is preserved.
Shortcut to Adiabatic (STA) techniques are devised for this purpose~\cite{GueryOdelin2019}. 
Among the others, counterdiabatic (CD) driving is a versatile technique in which by adding auxiliary drives to the system the diabatic transitions resulting from fast modification of the Hamiltonian are averted and the system can be driven along a specific instantaneous eigenstate, thus giving the outcome of an adiabatic process in much shorter times.

%===================================
\subsection{Counterdiabatic driving}
\label{sec:CDV}
Consider a time-dependent Hamiltonian $\hat{H}_0(t)$ with its instantaneous eigenstates $\ket{n(t)}$ satisfying the eigenvalue equation $\hat{H}_0(t) \ket{n(t)} = E_n(t) \ket{n(t)}$.
According to the quantum adiabatic theorem if the system is initially at any eigenstate it will remain in the same eigenstate when changing the Hamiltonian over time, provided that those changes are slow enough.
In contrast, when the process is fast diabatic transitions populate other instantaneous eigenstates.
In a transitionless process, such undesirable excitations in the system are compensated for by employing an auxiliary Hamiltonian $\hat{H}_1(t)$.
Therefore, the system ideally remains in its instantaneous eigenstate even if the adiabatic conditions are not satisfied.
It is straightforward to show that~\cite{Berry2009}
\begin{equation}
\hat{H}_1(t) = i\hbar \sum_{n\neq m} \frac{\proj{n(t)} \partial_t \hat{H}_0(t) \proj{m(t)}}{E_m(t) - E_n(t)}.
\label{eqn:HCD}
\end{equation}
Hence, dynamics of the system under the Hamiltonian $\hat{H}(t) = \hat{H}_0(t) + \hat{H}_1(t)$ gives the eigenstates of $\hat{H}_0(t)$ for arbitrary processing times, provided $\hat{H}_1(t_{\rm i}) = \hat{H}_1(t_{\rm f}) =0$.
This last condition ensures equality of the $\hat{H}_0(t)$ and $\hat{H}(t)$ eigenstates at the boundary times.
% To show this, we note that a unitary evolution operator $\hat{U}(t)$ satisfies the equation
% \begin{equation}
% 	i\hbar \partial_t \hat{U}(t) = \hat{H}(t) \, \hat{U}(t)
% \end{equation}
% thus the  transitionless Hamiltonian can be constructed using it
% \begin{equation}
% 	\label{eqn:H_from_U}
% 	\hat{H}(t) = i \hbar \left({\partial_t \hat{U}(t)} \right) \hat{U}^\dag(t) = i\hbar \dot{\hat{U}} \hat{U}^\dag.
% \end{equation}
% Choosing $\hat{U}$ as
% \begin{equation}
% 	\hat{U}(t) = \sum_n e^{i \xi_n(t)}  \ket{n(t)} \left< n(0) \right|
% \end{equation}
% and using \eqref{eqn:H_from_U} yields
% \begin{equation} 
% \begin{split}
% 	\hat{H}(t) &= \sum_n E_n(t) \left|n(t)\right>\left<n(t)\right| \\
% 		   &\quad + i\hbar \sum_n \left( \left|\partial_t n(t)\right>\left<n(t)\right| - \left<n|\partial_t n\right> \left| n(t)\right>\left<n(t)\right| \right) \\
% 		   &= \hat{H}_0(t) + \hat{H}_1(t).
% \end{split}
% \end{equation}
% This Hamiltonian drives an eigenstate $\ket{n(t)}$ ($n=0,1,...$) of $\hat{H}_0(t)$ without allowing transitions to other states.

%===================================
\subsection{Model}
\label{sec:model}
We consider a double-well potential as the system where its lowest eigenstates are a quantum superposition of two distinct states.
Particularly, The groundstate of a symmetric DW potential consists of the symmetric superposition of the groundstate of each well, while its first excited state is an antisymmetric superposition of the same states.
Hence, a balanced mixture of these two gives a classical state.
The energy difference of the two eigenstates $\delta_{10} \equiv (E_1 - E_0)/\hbar$ is proportional to the probability with which the particle tunnels from one well to the other.
That is, a higher barrier energy results in a more distinguishability, but at the same time smaller energy gap between the two lowest states of the DW.
%Such that for an infinite barrier one has $\delta_{10} \to 0$ as expected for two identical noninteracting systems.
Therefore, when approaching a DW potential the thermal excitations become growingly fast and fade out the superposition features of the state.

For a massive particle in a spatial DW potential occupation of the groundstate means that a massive spatial symmetric superposition state is realized.
Various forms of DW potentials can be envisaged, but in this work we put our focus on the case were the explicit form of the potential is $V(z)=-\frac{1}{2}\nu z^2 +\frac{1}{4}\beta z^4$.
This, in principle, can be realized by subjecting an intrinsic Duffing resonator to an inverse parabola external potential, see Ref.~\cite{Abdi2016}.
A possible physical realization of $\nu$ and $\beta$ is discussed in Sec.~\ref{sec:scheme}, but in this section we examine some properties of such DW potential.
The above potential is centered at $z=0$ and its minima lie at $z = \pm z_0 = \pm \sqrt{\nu/\beta}$.
In quantum regime, the eigenstate whose energy is less than that of the central barrier is either a symmetric or an antisymmetric superposition of states almost localized in each well provided the barrier is high enough.
% These localized states can be approximated by eigenstates of a harmonic potential with frequency $\omega_0 = \sqrt{2\nu/m}$ and minima at the location of the well's minima, i.e. $\pm x_0$.
The groundstate of such a potential is a spatially distinguishable superposition state resulted from delocalization of the particle.
Roughly speaking it can be understood as an even cat state resulting from the symmetric superposition of the two harmonic well groundstates.
%The anti-parabola part of the potential can be realized externally.

In order to adiabatically prepare a system in the groundstate of potential $V$ a control mechanism must be invoked.
Indeed, the total potential can be decomposed in two parts $V(z) = V_{\rm m}(z) +V_{\rm e}(z)$: 
(i) The intrinsic Duffing oscillator $V_{\rm m}(z)=\frac{1}{2} \tilde\nu z^2 + \frac{1}{4} \tilde\beta z^4$.
(ii) The external potential $V_{\rm e}(z) = -|\alpha_2|z^2 +\alpha_4 z^4$, where $\alpha_2$ is an external softening force that opposes and eventually overcomes the intrinsic Hook force.
A positive $\alpha_4>0$ can strengthen the trap---which is partly weakened by the anti-parabola---and thus enhance the controllability of the system.
Nevertheless, for a given $\nu = \tilde\nu -|\alpha_2|$ a larger $\beta = \tilde\beta +\alpha_4$ will result-in a smaller spatial spacing in the DW minima $2z_0$.
Therefore, one in principle must engineer the optimal values of $\alpha_2$ and $\alpha_4$ for having both better control over the system and two spatially distinguishable wells, hence, the groundstate could represent a cat state.
Here, for the sake of simplicity we propose to operate in a regime were $\alpha_4 \approx 0$.
Hence, $\alpha_2$ is the only control parameter that tunes the potential, which is varied by time as explained in Sec.~\ref{sec:protocol}.
In other words, by changing the external potential one can tune $\alpha_2(t)$ over time and shape the double-well potential.

Our goal in this work is to prepare a massive object in a quantum superposition of two distinguishable states.
Therefore, we consider a mechanical resonator as the object and set $\tilde\nu = m\omega^2$.
By introducing the dimensionless fine tuning parameter $\zeta$ with the following equation
\begin{equation}
	\label{eqn:alpha2}
	\alpha_2(t) = -(1+\zeta(t)) \frac{m\omega^2}{2},
\end{equation}
% \begin{equation}
% 	\label{eqn:nu}
% 	\nu(t) = 2 \left| \alpha_2(t) \right| - m \omega^2
% \end{equation}
% We consider a mechanical element (system) with effective mass $m$ subjected to a DW (DW) potential [see Fig. \ref{fig:setup_DW}] \cite{abdi2016}.
Hamiltonian of the mechanical system in the regime where the quadratic part of the potential is tunable while the quartic term remains fixed reads
\begin{equation}
	\hat{H}_{\rm DW}(t) = \frac{\hat{p}^2}{2m} - \frac{1}{2}\zeta(t) m \omega^2 \hat{z}^2 + \frac{\beta}{4} \hat{z}^4,
\label{eqn:mechamil}
\end{equation}
% consists of three terms: a kinetic term, an inverted quadratic term and an attractive quartic term
% \begin{equation}
% 	\label{eqn:H_dw}
% 	\hat{H}_{\rm DW}(t) = \frac{\hat{p}^2}{2m} - \frac{\nu(t)}{2} \hat{x}^2 + \frac{\beta}{4} \hat{x}^4
% \end{equation}
where now $\hat{z}$ and $\hat{p}$ are the position and momentum operators for the only degree of freedom of the mechanical resonator, satisfying the canonical commutation relation $\left[\hat{z}, \hat{p} \right] = i \hbar$.
Note that $\zeta = -1$ retrieves the intrinsic elastic Hamiltonian when the external potential is extinguished.
The Hamiltonian in \eqref{eqn:mechamil} represents a confined system, and thus, has a discrete spectrum which is bounded from below.
Therefore, the instantaneous eigenstates and eigenenergies satisfying the eigenvalue equation $\hat{H}_{\rm DW}(t)\ket{n(t)} = E_n(t)\ket{n(t)}$ are indexed as $n=0,1,2,3,\cdots$ with ascending order of the eigenvalues such that $E_0<E_1<E_2<\cdots$.
Furthermore, due to the symmetric nature of the DW Hamiltonian its eigenstates are simultaneous eigenvalues of the parity operator $\hat\Pi$ whose effect is $\hat\Pi \hat{z}\hat\Pi^\dag = -\hat{z}$ and $\hat\Pi \hat{p}\hat\Pi^\dag = -\hat{p}$.
The states with even indices $n=0,2,4,\cdots$ are found to have even parity, while the rest ($n=1,3,5,\cdots$) exhibit odd parity.

It is worth mentioning that one could also consider a scenario with two variable potential parameters where both $\zeta$ and $\beta$ are tunable.
Nevertheless, the it is usually easier and experimentally more feasible to deal with smaller number of variables.
Furthermore, parabolic (and anti-parabolic) potentials are easier to engineer.

\begin{figure}[tb]
\includegraphics[width=\columnwidth]{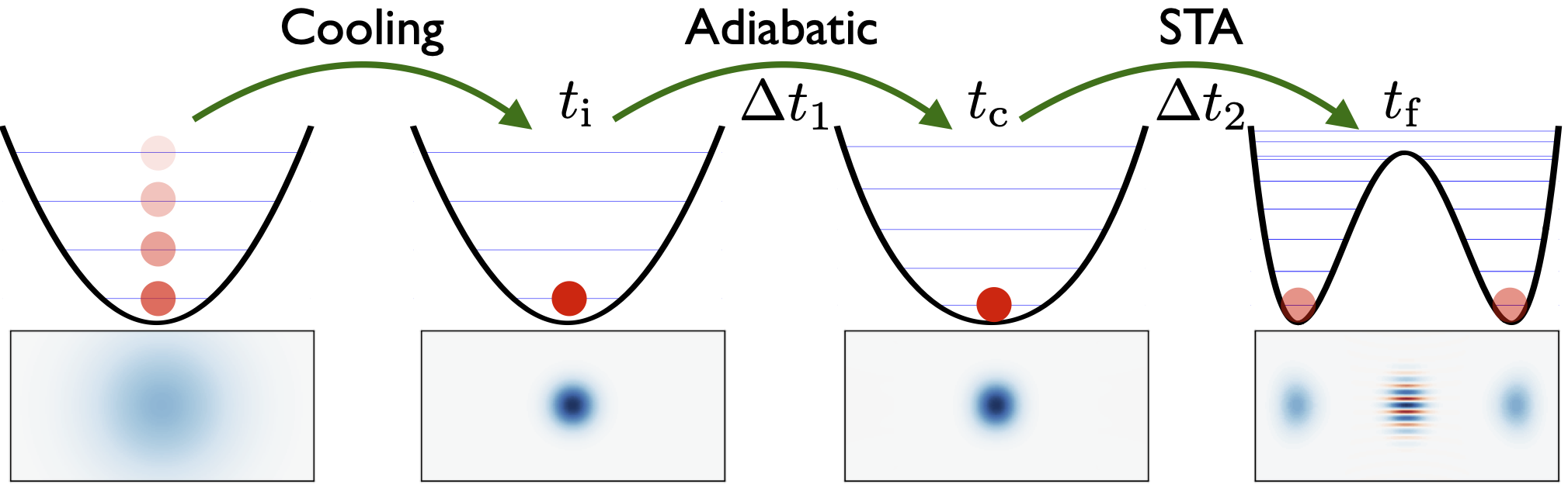}
\includegraphics[width=0.9\columnwidth]{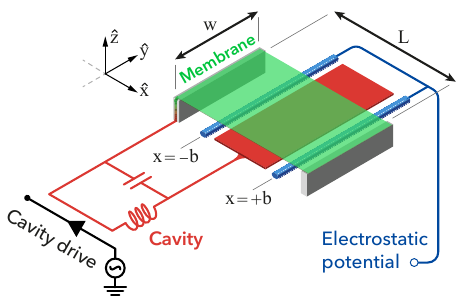}
\caption{Scheme of the proposed protocol (top panels) and the circuit quantum electromechanical setup (lower panel).
The microwave cavity capacitively couples to the graphene membrane (the green sheet) through the gate electrode (red rectangle).
The double-well potential forms by applying electrostatic forces via two parallel rod electrodes (blue lines).
}\label{fig:setup_DW}%
\end{figure}
%%%%%%%%%%%%%%%%%%%%%%

%===================================
\subsection{Protocol and dynamics}
\label{sec:protocol}
The goal is to achieve the ground state of $\hat{H}_{\rm DW}$ with $\zeta>0$ with a high fidelity despite its exposure to the environmental noises.
To this end, we propose a protocol where first the mechanical mode is cooled down to its ground state when it is in (an almost) harmonic trap ($\zeta(t_{\rm i})=-1$).
This can be performed by various techniques, e.g. sideband cooling through a coupled cavity mode~\cite{Marquardt2007, WilsonRae2007}.
Then the external potential is turned on and gradually increased until a DW forms ($\zeta(t_{\rm f})>0$).
However, the ramp function that takes the system from $\zeta(t_{\rm i})$ to $\zeta(t_{\rm f})$ is crucial.
Moreover, there is a trade off between the system thermalization and the diabatic excitations built-up in the system during the process.
For short protocol durations the Landau–Zener transitions are prominent, while for longer times the destructive effects of the environmental noise reduce the state fidelity.
Therefore, we propose to speed up the process and meanwhile compensate for the unwanted transitions in the system by employing a STA protocol based on the counterdiabatic driving.
A comprehensive CD scheme in a continuous variable system like the one studied in this work demands an infinite number of drives.
Nevertheless, in the next section we show that the purpose can still be fulfilled to a great extend by carefully selecting a few transitions.
This is specially crucial for the feasibility of our protocol as such drives can be experimentally implemented through the cavity modes, see Ref.~\cite{Abdi2016} for more details.
Therefore, a limited number of cavity modes are sufficient for attaining the superposition state with high fidelity.

Hence, the protocol for creating the cat state is the following three steps ($t_{\rm i} <t_{\rm c} <t_{\rm f}$):
(i) $t<t_{\rm i}$: Cooling the system to its groundstate when $\zeta(t) = -1$.
Indeed the harmonic nature of the system at this value of $\zeta$ allows one to employ a standard sideband cooling mechanism.
(ii) $t_{\rm i} \leq t \leq t_{\rm c}$: Turning on the external potential and \textit{adiabatically} approaching the buckling point.
That is, tuning the potential to a small but negative $\zeta(t_{\rm c})\lesssim 0$.
With a proper choice of $\zeta(t_{\rm c})$ the thermal excitation rates remain low and the system retains its instantaneous groundstate with a high fidelity for a reasonably slow ramp $\zeta(t_{\rm i}\to t_{\rm c})$.
(iii) $t_{\rm c} \leq t \leq t_{\rm f}$: Employing CD drives for a few lowest transitions while the potential is quickly modified to $\zeta(t_{\rm f})=\zeta_{\rm f}>0$, see Fig.~\ref{fig:setup_DW} for a schematic presentation.
In fact, the symmetric nature of the potential in our work demands for the conservation of parity.
Therefore, no diabatic transition occurs from the groundstate to the first excited state or any other eigenstate with odd parity.
Hence, one only needs to compensate for the diabatic transitions to the higher symmetric energy levels.
As one can infer from Eqs.~\eqref{eqn:HCD} and \eqref{eqn:mechamil} the counterdiabatic drive is $\propto \hat{z}^2$ which produces these desired transitions.

To study dynamics of our open quantum system, we employ the master equation formalism.
However, notice that the extreme nonlinearity of the system when the external softening force is comparable to the intrinsic stiffness demands for a careful open quantum system treatment.
Specially when the system enters the DW regime.
In Ref.~\cite{Abdi2016} some of us derive the proper dissipators that constitute the master equation describing the interaction of the system with its surrounding environment as the following 
\begin{align}
\partial_t\rho(t) &= \frac{1}{i\hbar} \big[ \hat{H}_{\rm DW}(t)+\hat{H}_{\rm drv}(t), \rho(t) \big] \nonumber\\
&+ \frac{1}{2}\big[ \hat{z},\rho(t)\hat{A}(t)^\dag - \hat{A}(t)\rho(t) \big],
\label{eqn:master}
\end{align}
where the coherent evolution includes both the double-well and drive Hamiltonians, where the latter intends to apply the counterdiabatic drive.
The incoherent part of the dynamics is described by the second term where we have introduced the jump operator $\hat{A} = \sum_{m>n} \gamma_{mn} \big( \bar{N}(\delta_{mn}) \tran{m}{n}  +  \left[ \bar{N} (\delta_{mn})+1 \right] \tran{n}{m} \big)$ in which $\gamma_{mn} = \big(2m\omega\delta_{mn}/\hbar Q\big)\langle m|\hat{z}|n\rangle$ is the decay rate from state $\left|m\right>$ to $\left|n\right>$.
Here, $\delta_{mn} = (E_m - E_n )/\hbar$ is the transition frequency, $Q$ is the quality factor of harmonic mechanical oscillations, and $\Nbos(\Omega)=[\text{exp}(\hbar \Omega / k_{\rm B} T) - 1]^{-1}$ is the occupation number at the temperature $T$, where $k_{\rm B}$ is the Boltzmann constant.
Notice that in the above definitions $E_n$ and $\ket{n}$ are the instantaneous eigenvalues and eigenstates of $\hat{H}_{\rm DW}$ and we have dropped their explicit time dependence for better readability.
 
We numerically solve the above equation with a time-dependent Hamiltonian through $\zeta(t)$ and with appropriate CD drive auxiliary Hamiltonian~\cite{Johansson2013}.
Each step of the protocol is performed separately and outcome of the previous step is fed as the initial state for the next one, see Sec.~\ref{sec:results} for the details.

%========================================
\subsection{Setup}
\label{sec:scheme}
Here, we consider and discuss an experimental setup as a possible implementation of the above discussed scheme.
A rectangular monolayer of graphene with dimensions $w\times L$ is employed as the mechanical resonator, where the goal is to establish a superposition of deflections in two directions perpendicular to its surface.
Thanks to the large Young modulus, flexural modes of a free-standing graphene membrane experience large Duffing nonlinearity making them a good candidate for implementing our protocol [Appendix~\ref{app:elastic}].
The polarizability of graphene allows us to apply the external softening force by applying an electrostatic potential~\cite{Hwang2007}.
Two line electrodes can provide the anti-parabola while maintaining symmetry properties of the membrane, see Fig.~\ref{fig:setup_DW} for an illustration and Appendix~\ref{app:electrostatic} for the details.

When the pinned membrane boundary conditions are applied, one has $m=\frac{1}{2}\varrho L w h$ and $\beta = Y h w/(8\pi^4 L^3)$ for the effective mass and Duffing nonlinearity of the resonator, respectively, while the mode frequency is mostly determined by the tensile force, see Appendix~\ref{app:elastic}.
Here, $\varrho = 2.26\times 10^{3}$~kg/m$^3$ and $Y = 1.02$~TPa are the graphene bulk mass density and Young modulus.
By considering a monolayer graphene with $\{L, w, h\} = \{5,1, 3.35\times 10^{-4}\}~\mu$m one finds $m = 1.9 \times 10^{-12}$~kg and $\beta = 3.3 \times 10^{13}$~J/m$^4$.
We assume a mechanical frequency of $\omega/2\pi = 2$~MHz for the fundamental flexural mode, the mode which has the highest coupling strength to the cavity.

In fact, to perform the initial cooling as well as for the CD driving a well-controlled quantum system is required to couple and interact with the membrane.
Therefore, we consider a circuit electromechanical system where a superconducting microwave cavity capacitively couples to the graphene membrane, see Fig.~\ref{fig:setup_DW} and Ref.~\cite{Song2014}.
The highly nonlinear nature of the double-well Hamiltonian in our scheme allows one to selectively stimulate the mechanical transitions that are necessary for producing the counterdiabatic Hamiltonian~\cite{Abdi2016}.

% with dimensions $a$ and $b$ in $x$ and $y$ directions, respectively and with effective mass $m$ subjected to a DW (DW) potential [see Fig. \ref{fig:setup_DW}]. The system is quite similar to one considered in ref. \cite{abdi2016}, but instead of a disk shape membrane we consider a rectangular doubly clamped membrane. In addition, the geometry of the electrodes which apply the electrostatic potential is different, so that we believe it will be easier to implement.

%§§§§§§§§§§§§§§§§§§§§§§§§§§§§§§§§§§§§
\section{Results}
\label{sec:results}
% We considered a membrane of LDG with a radius of $a \approx 1\mu m$ oscillating in its fundamental mode with frequency $\omega/2\pi \approx 26 \,$MHz and
% \begin{equation}
% 	\beta \approx 3.8 \times \frac{Y h}{a^{2}} \approx 5.7 \times 10^{15} \, \text{J/m}^4
% \end{equation}
% where $Y$ is the Young modulus of LDG and $h=0.34 \, \text{nm}$ is the thickness of the membrane [see Ref. \cite{abdi2016}].
The eigenstates of the Hamiltonian \eqref{eqn:mechamil} are numerically computed for different values of $\zeta$.
By examining different values of $\zeta >-1$, one clearly sees that the potential changes from an almost harmonic trap at $\zeta=-1$ to highly nonlinear single-well trap for $\zeta \lesssim 0$ and sets to form a DW shape when $\zeta>0$ [see Appendix~\ref{app:nuermical}].
A large positive $\zeta$ gives a deep DW with several pairs of closely spaced energy levels with symmetric and antisymmetric wave functions.
Even though such values of $\zeta$ provide a larger spatial separation in the components of the superposition state, achieving their groundstate becomes increasingly difficult as $\zeta$ increases.
Indeed, the effect is twofold:
First, the first excited state which is the antisymmetric superposition of the up and down deflections becomes easily accessible by thermal excitations.
This can quickly result in a thermal mixture of the two lowest states which is a classical state.
Second, a tight set of energy levels leads to more complicated diabatic transition and thus an exhaustive counterdiabatic driving scheme must be invoked, which in turn demand more experimental resources.
Therefore, here we consider the modest value of $\zeta_{\rm f} = +3 \times 10^{-4}$.
The optimal value of the intermediate $\zeta$ that the third stage of the protocol starts is numerically found to be $\zeta_{\rm c} = -2.5 \times 10^{-4}$.

% Since we're interested in the ground state $\left|0\right>$ we start to add an approximate  version of \eqref{eqn:HCD} to the original Hamiltonian after $\zeta = -2.5\text{e-}4$ and drive $\left|0(t)\right>$ under
Next we notice that $\partial_t\hat{H}_{\rm DW} \propto \hat{z}^2$.
Therefore, the diabatic transitions can only happen among the even and odd subspaces.
Consequently, the counterdiabatic Hamiltonian can be divided into two terms consisting transitions among the even and odd subspaces, $\hat{H}_{\rm drv} = \hat{H}_{\rm drv}^+ +\hat{H}_{\rm drv}^-$.
Since in our protocol the system is initially in the groundstate $\ket{0}$ with a fidelity close to the unity and the goal is to keep the system in its instantaneous groundstate, it is enough to only include counterdiabatic transitions in the even subspace.
Hence, we discard the odd subspace drive terms and get
\begin{equation}
\hat{H}_{\rm drv}(t) =  -\half i\dot{\zeta}m\omega^2 \sum_{\substack{m > n \\ n\in \mathbb{E}}} \tfrac{\bra{n(t)}\hat{z}^2\ket{m(t)}}{\delta_{mn}(t)}\tran{n(t)}{m(t)} +\text{H.c.},
\label{eqn:HCD_even}
\end{equation}
where $\dot\zeta$ is the time derivative of the control parameter.
%\begin{equation}
%\label{eqn:HCD_tilde}
%\begin{split}
%	\hat{\tilde{H}}_\text{CD}(t) =  i(\frac{-3\hbar}{2}) \dot{\zeta} &\zeta^{-5/2} \bigg[ \frac{\left|0\right> \left<0\right| \hat{A} \left| 2\right> \left< 2 \right| - c.c.}{E_2 - E_0} \\
%	& \quad\,\,\,\,+ \frac{\left|0\right> \left<0\right| \hat{A} \left| 4\right> \left< 4 \right| - c.c.}{E_4 - E_0}\bigg]
%\end{split}
%\end{equation}
Despite the complicated form of the above Hamiltonian we will show below in Sec.~\ref{sec:sta} that considering and stimulating a few transitions is enough for attaining a high fidelity groundstate of the double-well Hamiltonian.
This also proves the experimental feasibility of our protocol.
To analyze the performance of our protocol in different situations we compute the fidelity $F=\tr\{\sqrt{\sqrt{\rho}\sigma\sqrt{\rho}}\}$ of the outcome state $\rho$ at the end of the protocol that results from the full dynamics of the master equation \eqref{eqn:master} with the target state $\sigma$ which is the groundstate of $\hat{H}_{\rm DW}$ for $\zeta=\zeta_{\rm f}$.
In Fig.~\ref{fig:adiabatic}(a) the Wigner function of the target state is presented.

%========================================
\subsection{Full adiabatic preparation}
\label{sec:full_adiabatic_prep}
\begin{figure}[tb]
\includegraphics[width=\columnwidth]{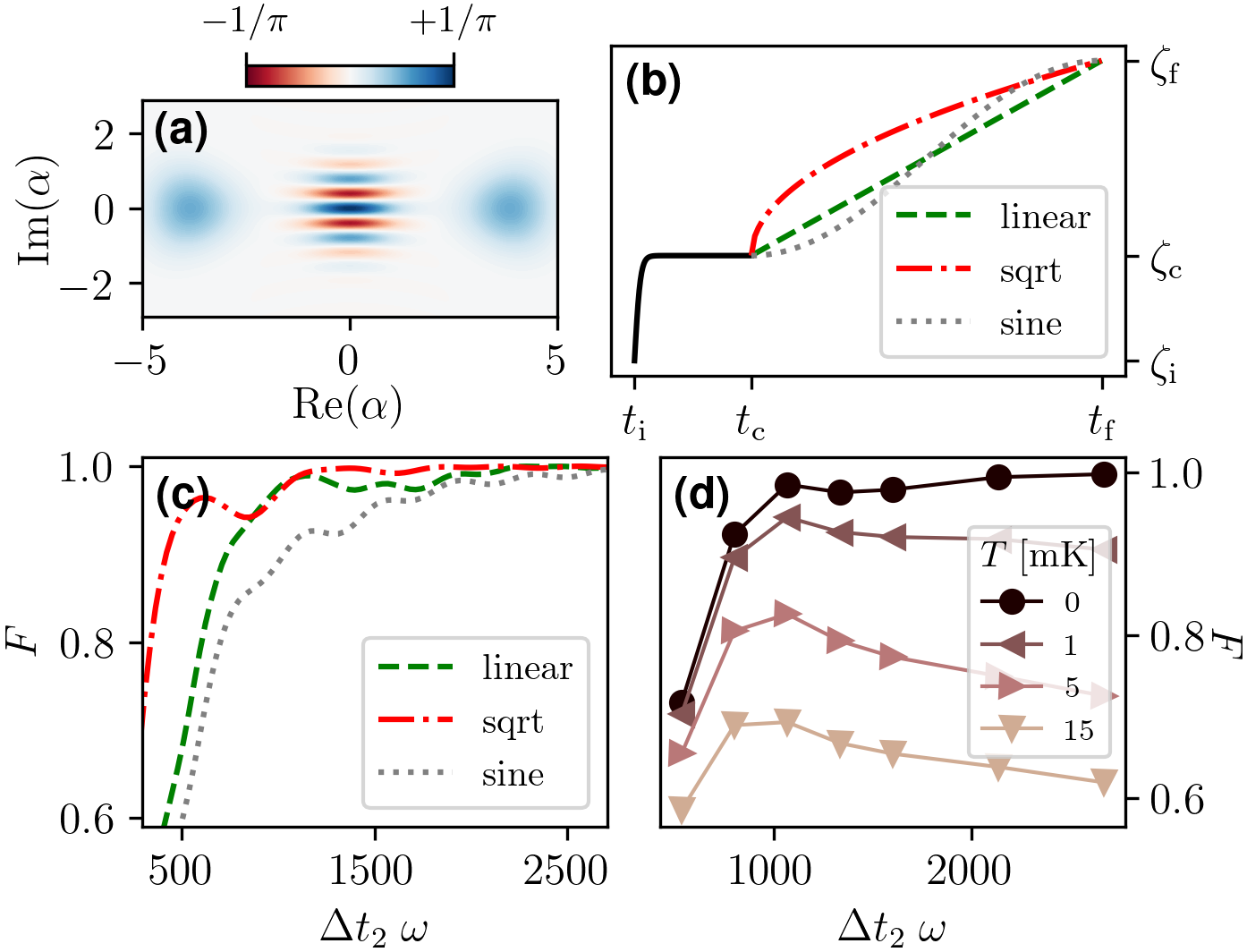}
\caption{%
(a) Wigner function of groundstate of the double-well potential with $\zeta=3\times 10^{-4}$, the target state.
(b) The ramp functions employed in the second (solid black line) and third (various line styles shown in the legend) steps of the protocol. The values of $[t_{\rm i},t_{\rm c},t_{\rm f}]$ and $[\zeta_{\rm i},\zeta_{\rm c},\zeta_{\rm f}]$ are not to the scale for a better illustration.
Fidelity of the outcome state $F$ with respect to the target state versus duration of the last step of the protocol $\Delta t_2$:
(c) For an isolated system when three different ramp functions are considered.
(d) At different temperatures when a linear ramp function is employed.
Note that the green dashed line in (c) and the black circles in (d) correspond to the same parameters and conditions.
}%
\label{fig:adiabatic}
\end{figure}

We first consider the case where no CD transition drives are applied during the state preparation.
The only difference with the protocol described above is that in all steps $\hat{H}_{\rm drv}=0$.
Moreover, we assume a perfect initialization of the system where the harmonic oscillator ($\zeta = -1$) is cooled down to its ground state with a fidelity of unity.
The effect of non-ideal initialization will be studied later, see Sec.~\ref{ssec:imperfections}.
In the second step of the protocol the potential applied to the electrodes is tuned such that the control parameter changes from $\zeta_{\rm i}=-1$ to $\zeta_\text{c}=-2.5\times 10^{-4}$.
Our numerical calculations show that this stage can be performed in a reasonably short time interval with almost no state degradation, neither due to thermalization nor from diabatic transitions~\cite{Aslani2023}.
Employing a proper ramp function, however, is necessary.
For $\zeta<0$ the energy level spacing scaling is upper-bounded by $\delta_{mn} \gtrsim \omega|\zeta|^{1/2}$.
Hence, a ramp adjusted with the pace of gap closing rate can keep the system dynamics away from any diabatic transition to arbitrary negative values of $\zeta$, provided that $\Delta t_1 \sim 1/\omega$, see the solid black line in Fig.~\ref{fig:adiabatic}(b).
The third step of the protocol is the same as the previous one but the ramp function for evolving the system into a double-well potential needs to be adjusted properly.
The energy level differences exhibits a complex behavior at the point the double-well potential sets to form ($\zeta \gtrsim 10^{-4}$).
Extremely complicating pattern of the diabatic transitions.
Nonetheless, for the sake of clarity and simplicity, here we consider and compare the performance of simple ramp functions.
It is worth mentioning that one could also employ the more advanced fast quasiadiabatic approach where the ramp functions are systemically optimzed~\cite{Garaot2015}.

To study performance of the fully adiabatic protocol we consider three different ramp functions for the third step of the protocol:
Linear evolution of $\zeta$ over time, square root (sqrt), and sinusoidal (sine) function [see Fig.~\ref{fig:adiabatic}(b)].
The smooth sine function satisfies the initial and final time requirements of CD drives in the STA protocol, and thus, is suitable for comparing performances of the two protocols.
The results for the outcome state show that when there is no thermal noise ($T=0$) a high fidelity final state can be achieved for long enough processing time through either of the ramp functions.
This is clear from Fig.~\ref{fig:adiabatic}(c) where the fidelity of the outcome state with respect to the target state is plotted versus $\Delta t_2$.
In the next step that effect of the thermal noise is studied we only consider the simple linear ramp function.
Now we include the thermal noise effect in the computations.
As expected even for very low temperatures the noise is largely detrimental.
Fig.~\ref{fig:adiabatic}(d) shows that the fidelity rapidly decreases as the thermal noise is introduced to the system.
In obtaining these results we have set $\Delta t_1=1/\omega$ for the second step of the protocol.
%%%%%%%%%%%%%%%%%%%%%%%%
%%%%%%%%%%%%%%%%%%%%%%%%

%========================================
\subsection{Shortcut to adiabaticity}
\label{sec:sta}
The counterdiabatic transition drives are now included in the third step of the protocl.
%We begin our analysis with the simplest scenario by setting $n,m\in\{0,2\}$ in Eq.~\eqref{eqn:HCD_even}.
Let us emphasize that the level spacing for $\zeta_{\rm i}\leq\zeta\leq\zeta_{\rm c}$ values are large enough to allow for the fast evolution of the system without state degradation through diabatic transition or the thermal noise in the temperatures studied in this work.
Therefore, the CD drives are only employed for the last step of the protocol where the system enters the DW realm.
As above we set $\Delta t_1 = 1/\omega$ and find a high fidelity for the state at $t=t_{\rm c}$.
In the third part of the protocol, the counterdiabatic drives producing the Hamiltonian \eqref{eqn:HCD_even} with $n,m\in\{0,2,\cdots\}$ are introduced, while the control parameter takes the smooth ramp that connects $\zeta_{\rm c}$ to $\zeta_{\rm f}$ in the time duration of $\Delta t_2$.
The ramp function utilized in this part takes the form of a quarter-sine function, as illustrated in Fig.\ref{fig:adiabatic}(b). This choice ensures the fulfillment of the boundary conditions outlined below Eq.\eqref{eqn:HCD}, which are imperative for a STA protocol.
%The first one, nonetheless, is optimized numerically such that when the system is in almost a Harmonic potential is swiftly brought to the anharmonic regime at which point the control parameter slows down to avoid diabatic transitions.
The final outcome of the protocol is a cat state whose Wigner function is very close to the one presented in Fig.~\ref{fig:adiabatic}(a).
The outcome has almost perfect match with the groundstate of the DW potential with the fidelity $F \approx 99.7 \% $ at the finite dilution refrigerator temperatures.
Achieving the ground state of a deep DW potential with a fidelity as high as $99 \%$ ensures that the system is in a very non-classical state.
This is clearly visible from the large spatial separation of the coherent lobes and negative features of the Wigner function.
%%%%%%%%%%%%%%%%%%%%%%%%
\begin{figure}[tb]
\includegraphics[width=\columnwidth]{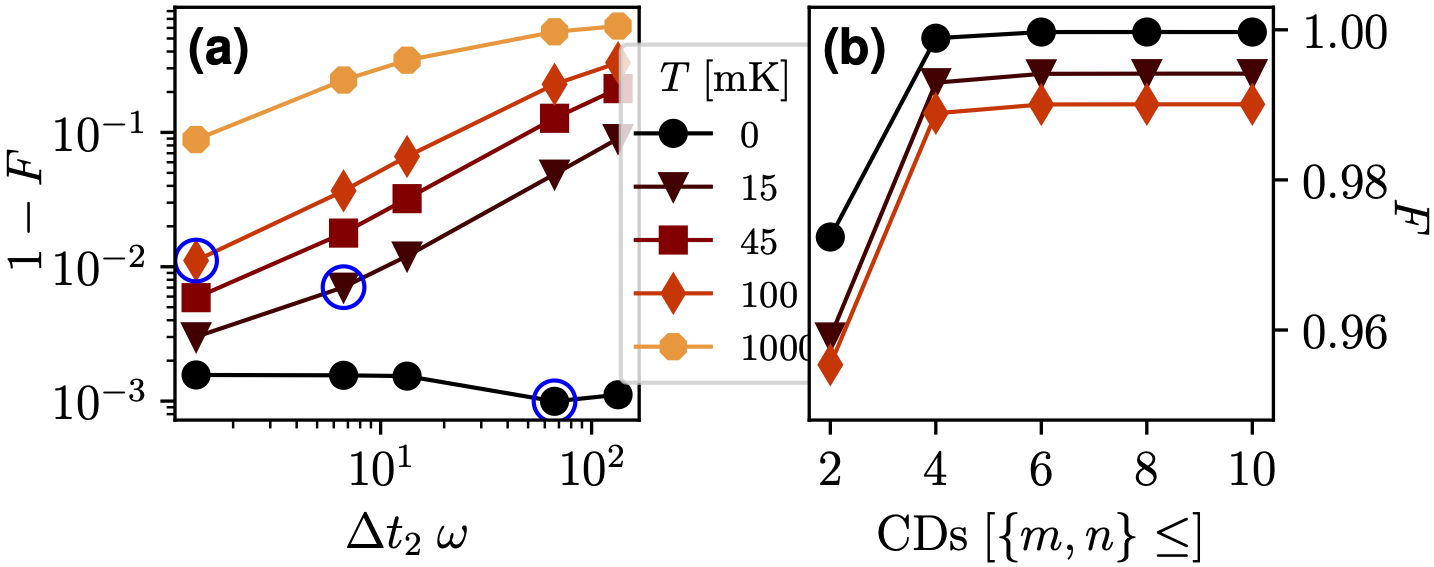}
\caption{(a) Infidelity ($1-F$) of the final state versus duration of the last step of the protocol $\Delta t_2$ for different ambient temperatures. Here, only transition up to fourth level are taken in the numerics, i.e. $m,n \in \{0,2,4\}$ in Eq.(5).
In (b) the effect of including different sets of CD drives in the fidelity of the final state $F$ is presented for three specific cases marked by blue circles in (a). The numbers on the horizontal axis refer to the highest level considered, e.g. $6$ stands for $\{m,n\}=\{0,2,4,6\}$.
}
\label{fig:sta}
\end{figure}
%%%%%%%%%%%%%%%%%%%%%%%%

Effect of the environment temperature $T$ that determines the thermal decoherence rate as well as the ramp function on the fidelity are investigated.
The results are shown in Fig.~\ref{fig:sta}(a), where the infidelity $1-F$ of the final state is plotted against duration of the second ramp $\Delta t_2$ at different temperatures.
The destructive effect of ambient temperature is clear from the curves.
Furthermore, a faster ramp accompanied with proper CD drives can tremendously reduces such destructive effects.
The thermal noise is very prohibitive as it can be inferred by comparing the results in the absence of thermal noise (black circles) with the ones in its presence.
For $T=0$~K the error slightly reduces as the duration time increases.
%This trend is similar to the results obtained for the adiabatic passage, though with a much faster pace.
At finite temperatures the error gets about a hundred times larger when the duration time $\Delta t_2$ increases from $0.1$ to $10~\mu$s.
This, however, seems to be partially saturated as the temperature increases.
In these computations only CD drives up to the fourth level are taken into account.
We shortly provide numerical evidence that this is indeed enough for attaining a high fidelity at the end of the protocol.

We now study effect of the number of levels included in the counterdiabatic transition on the fidelity of the outcome state.
For this, in Eq.~\eqref{eqn:HCD_even} first only the lowest transition, i.e. $\ket{0}\leftrightarrow \ket{2}$, is considered.
We find that even though compensating for this transition significantly improves the results, it still leaves room for further enhancement.
The matrix element $\bra{4}\hat{z}^2\ket{0}$ assumes a rather large value and thus its diabatic transitions are appreciable.
Therefore, introducing a deexciting mechanism for this transition through the CD drives is expected to enhance the final fidelity.
This indeed is numerically confirmed to be the case as the fidelity takes a leap in Fig.~\ref{fig:sta}(b).
The enhancement in the final fidelity resulting from the inclusion of higher transitions is only incremental and seems to be saturating.
Hence, one only requires a limited number of counterdiabatic drives, only three, to attain a high fidelity cat state.
This is a crucial achievement that matches properties of our proposed setup for the experimental implementation.
Indeed, this means by employing three cavity modes and driving them with appropriate detuning and amplitude a high fidelity macroscopic superposition state can be attained.

The significance of the STA protocol can indeed be deduced by contrasting the black dots in Fig.~\ref{fig:sta}(a) to the curves in Fig.~\ref{fig:adiabatic}(c) and (d).
For an isolated system the high fidelity (small infidelity) state preparation is attained in much shorter $\Delta t_2$ durations (about two orders of magnitude smaller) when the STA protocol is employed.
Without CD drives and for short duration of the third step of the protocol, the fidelity of the outcome state is very small.
For example, for $\Delta t_2 = 500/\omega$ and $T=0$ the fidelity is about $F\approx 60\%$ for a sinusoidal ramp, see the leftmost part of the dotted curve in Fig.~\ref{fig:adiabatic}(c).
This is far less when shorter $\Delta t_2$ values are employed.
We have performed the computations for $\Delta t_2=110/\omega$ (the longest duration considered for STA protocol) and $T=0$ and find a fidelity of $F \approx 21\%$ which is much smaller than the value obtained with the STA protocol ($F\approx 99.9\%$).

%===================================
\subsection{Imperfections}
\label{ssec:imperfections}
\begin{figure}[tb]
\includegraphics[width=0.51\columnwidth]{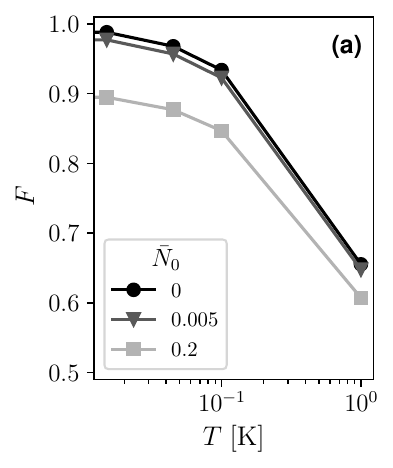}
\includegraphics[width=0.46\columnwidth]{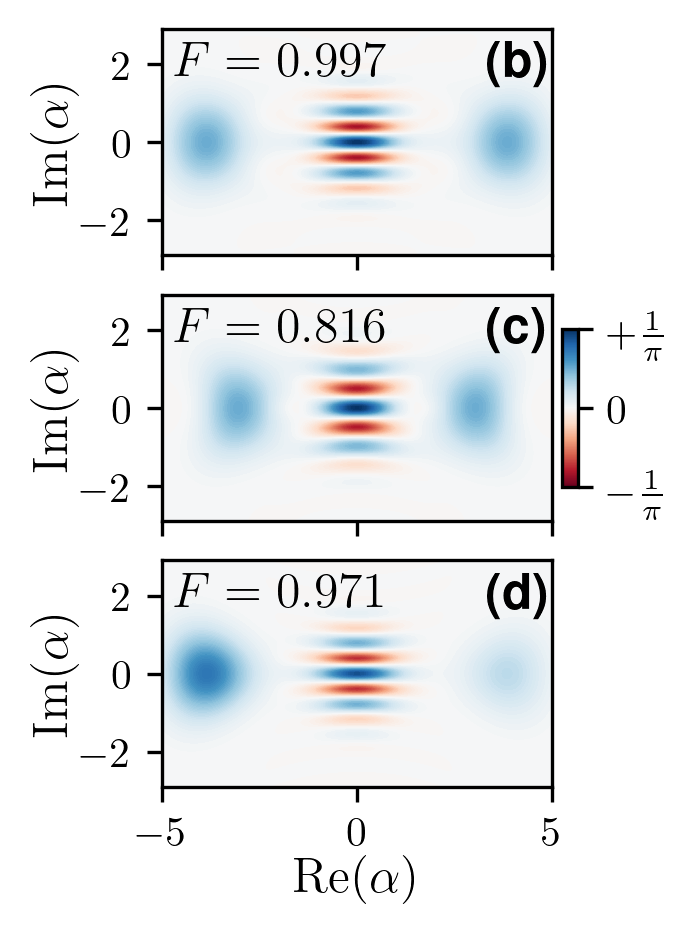}
\caption{Effect of imperfections in the fidelity of the outcome state:
(a) The fidelity as a function of ambient temperature for three different initial occupation numbers of the harmonic system $\Nbos_0$.
(b) and (c) Wigner functions of the outcome state of the protocol when the DW potential is symmetric ($\xi=0$) and asymmetric ($\xi=0.01$), respectively.
In (d) the Wigner function for the groundstate of an asymmetric DW is shown.
The numbers in the corner of the plots indicate their fidelity with respect to the target state.
Here, $\Delta t_2 = 0.1~\mu$s, and CD drives up to the fourth level are employed.
In (b)--(d) the ambient temperature is set to $T=15$~mK.
}
\label{fig:imperf}
\end{figure}

Alongside the thermal noise that has been considered in the above numerical analyses, there are several other effects that can still hinder achievement of a high fidelity cat state.
Here, we consider two most prominent effects, namely the case of non-ideal initial state and the asymmetry of the double-well potential.
Regarding the former, despite successful experimental results, a perfect cooling of the system to its groundstate is not attainable and the residual thermal occupations can be prohibitive for the final cat state.
Therefore, instead of a pure groundstate in the first step of the protocol one indeed must consider a thermal state, though with very low thermal occupation numbers, as the input for the second step of the protocol.
We numerically analyze this effect for two different initial occupation numbers $\Nbos_0 = \{5\times 10^{-3},0.2\}$ that correspond to finding the initial harmonic system in its groundstate with the probabilities $\approx\{99\%,90\%\}$.
The results suggest that although for the low ambient temperatures the fidelity of the final state decreases with almost the same proportion, the effect is overwhelmed at higher temperatures with the thermal noise, see Fig.~\ref{fig:imperf}(a). 

Next we study the other obstacle in attaining the cat state, the asymmetry in the DW potential.
The small yet non-vanishing $z^3$ contribution in the electrostatic potential introduces asymmetries in the total potential that the mechanical resonator `feels', see Appendix~\ref{app:electrostatic}.
This breaks the parity symmetry of the system and as a result alters the counterdiabatic transitions.
To see to what extend such asymmetries can affect outcome of our protocol we add the extra term $\frac{1}{3}\xi \hat{z}^3$ to the Hamiltonian \eqref{eqn:mechamil} and perform the numerical computations.
Interestingly, our results show that our protocol is robust against such asymmetric terms in the potential.
That is, by only considering CD driving among the even states and more specifically for only $n,m\in \{0,2,4\}$ in Eq.~\eqref{eqn:HCD_even} the outcome state is a cat state with two identical coherent lobes which resembles the groundstate of a symmetric DW.
The fidelity values confirm this visual assesment.
The protocol outcome state has an overlap of $F\approx 81.6\%$ with the symmetric DW groundstate, which is higher than that of the asymmetric case ($\approx 79.3\%$).
In Fig.~\ref{fig:imperf}(b)--(d) Wigner functions of the three following states are shown respectively:
the protocol outcome when the potential is symmetric ($\xi=0$),
the outcome state for an asymmetric DW with $\xi = 0.01$,
and groundstate of the asymmetric DW ($\xi = 0.01$).
The fidelities of these states with respect to the target state are given in each plot.
%%%%%%%%%%%%%%%%%%%%%%%%
%%%%%%%%%%%%%%%%%%%%%%%%

%§§§§§§§§§§§§§§§§§§§§§§§§§§§§§§§§§§§§
\section{Readout}
\label{sec:tomography}
Mechanical resonators in a quantum state are highly sensitive to the decohering effects when subject to a direct classical transduction.
Therefore, to gather information about their state an indirect readout mechanism through a well-controlled quantum system must be invoked.
In this path, yet two different directions can be taken:
To perform a full state tomography, which is usually very demanding given the inaccessibility of the mechanical resonators and the huge number of measurements that are necessary to reproduce the Wigner function, see e.g.~\cite{Muhonen2019, Enzian2021}.
Alternatively, one could search for exclusive traces that signal essential properties of the quantum state of interest.
Specifically, traces that are converted into the properties of an interacting quantum system.
Analyzing the spectrum of an outgoing cavity mode interacting with the mechanical mode is an experimentally feasible approach~\cite{Kippenberg2007, Groblacher2009, Clerk2010, Aspelmeyer2014}.
This method has been used to verify groundstate preparation of a harmonic mechanical resonator~\cite{Rocheleau2010, Chan2011, Teufel2011}. 

The highly nonlinear nature of the DW potential gives rise to an anharmonic mechanical spectrum.
In this section we take advantage of this property and propose a method to read out the mechanical state for verifying the prepared cat state via spectroscopy of a driven cavity field coupled to the mechanical mode.
A cavity mode weakly coupled to the mechanical resonator adiabatically follows its dynamics and carries the information therein~\cite{Vitali2007, Hartmann2008, Rips2012, Abdi2015b}.
Thus, the outgoing field of the cavity mode has fingerprints of different mechanical transitions weighted by the occupation of the mechanical levels.
To study the effect we include the cavity mode and its interaction with the mechanical mode in the Hamiltonian.
The free dynamics of a cavity mode with frequency $\omega_{\rm c}$ driven at frequency $\omega_{\rm d}$ in the frame rotating at the drive frequency is described by the following Hamiltonian
\begin{equation}
	\hat{H}_{\rm c} = -\hbar\Delta_0\hatd{a}\hat{a} +i\hbar\mathcal{E}_{\rm c}(\hatd{a} -\hat{a}),
\end{equation}
with the detuning $\Delta_0 = \omega_{\rm d} -\omega_{\rm c}$ and the drive amplitude $\mathcal{E}_{\rm c}$ which proportional to the input power, cavity frequency $\omega_{\rm c}$ and decay rate $\kappa$~\cite{Abdi2011}.
In a superconducting circuit electromechanical system the cavity capacitively couples to the flexural vibrations of the membrane, see Fig.~\ref{fig:setup_DW}.
The interaction Hamiltonian thus reads
\begin{equation}
	\hat{H}_{\rm int} = \hbar g_0\hatd{a}\hat{a}\hat{z},
\end{equation}
where $g_0$ is the single photon coupling strength.
This bare coupling is typically very small and thus is enhanced by strongly pumping the cavity.
In this regime, a linearized approximation for the interaction Hamiltonian adequately describes the system dynamics.
Hence, the total optomechanical Hamiltonian reads
\begin{equation}
	\hat{H}_{\rm tot} = -\hbar\Delta \hatd{a}\hat{a} +\hat{H}_{\rm DW} +\hbar g \hat{z}(\hat{a} +\hatd{a}),
\label{Hopto}
\end{equation}
where $\Delta$ is the modified detuning and $g$ is the drive-enhanced coupling strength.

To find the cavity spectrum we now study the cavity mode dynamics.
For this purpose it is convenient to work in the Heisenberg picture.
Dynamics of the cavity mode when driven in resonance ($\Delta=0$) is described by the following Langevin equation
\begin{equation}
	\dot{\hat{a}} = -\half\kappa \hat{a} -ig\hat{z} +\sqrt{\kappa}\hat{a}_{\rm in},
\label{langevin}
\end{equation}
where $\hat{a}_{\rm in}$ is the vacuum noise with the correlation function $\mean{\hat{a}_{\rm in}(t)\hat{a}_{\rm in}^\dag(t')}=\delta(t-t')$ and its all the other correlators vanishing.
The rigorous solution to Eq.~\eqref{langevin} is
\begin{equation}
	\hat{a}(t) = \int_0^t\! ds~ e^{-\frac{\kappa}{2}(t-s)}\big[-ig\hat{z}(s) +\sqrt{\kappa}\hat{a}_{\rm in}(s) \big],
\end{equation}
where we have dropped a transient term which is irrelevant in the steady-state limit that we are interested in.
Indeed, for a weakly coupled cavity the kernel in the integrand decays much faster than any other oscillation and effectively the higher limit of the integrand can be set to infinity.
Hence, by integration we arrive at
\begin{equation}
\hat{a}(t) \approx -ig\sum_{m,n}\frac{z_{mn}(t)\tran{m}{n}}{\kappa/2-i\delta_{mn}} +\frac{2}{\sqrt{\kappa}}\hat{a}_{\rm in}(t),
\end{equation}
where $z_{mn} = \bra{m}\hat{z}\ket{n}$ are the position matrix elements.
It is clear from the above equation that the cavity field inherits the mechanical state properties.
From the input-output theory the microwave field leaving the cavity ($\hat{a}^{\rm out} =\hat{a}^{\rm in} -\sqrt{\kappa}\hat{a}$) is thus carrying information about the mechanical transitions.
%\begin{equation}
%\hat{a}_{\rm out}(t) \approx -ig\sqrt{\kappa}\sum_{m,n}\frac{z_{mn}\tran{m}{n}}{\kappa/2-i\delta_{mn}} +\text{H.c.} -\hat{a}_{\rm in}(t),
%\end{equation}
Therefore, the output field spectrum can be exploited for readout of the mechanical state.
The steady-state spectrum of the outgoing cavity field can be computed by employing the quantum regression theorem~\cite{Carmichael2010}
\begin{equation}
	S(\Omega) = \sum_{n,m}\frac{\kappa g^2 z_{mn}^2}{\kappa^2/4 +\delta_{mn}^2}L_{mn}(\Omega)\rho_{nn},
\end{equation}
where $\rho_{nn}$ gives the $n$th diagonal element of the mechanical density matrix and we have introduced the Lorentzian function $L_{mn}(\Omega)=\frac{1}{\pi}\frac{\Gamma_{mn}}{(\Omega  -\delta_{mn})^2 +\Gamma_{mn}^2}$.
Here $\Gamma_{mn}$ is the decoherence rate of each mechanical transition whose value for $m<n$ is $\gamma_{mn}\Nbos(\delta_{mn})$, while for $m>n$ gives $\gamma_{mn}[\Nbos(\delta_{mn})+1]$.
In deriving the above equation, we have exploited the fact that $\delta_{mn}=-\delta_{nm}$ and because of the Hermitian nature of $\hat{z}$ one has $z_{mn}=z_{nm}^*$.
Note that the frequency of the cavity is set as the reference in above equation.
That is $\Omega=0$ corresponds to the cavity resonance frequency.
Apart from the main peak resulting from the drive at $\Omega=0$ state of the mechanical resonator in DW potential and its interaction with the cavity leaves its trace as sideband at the mechanical transitions $\delta_{mn}$ in the output spectrum.
Crucially, the fast decay rate of the cavity allows one to extract the mechanical information before its thermalization.

\begin{figure}[tb]
\includegraphics[height=3.5cm]{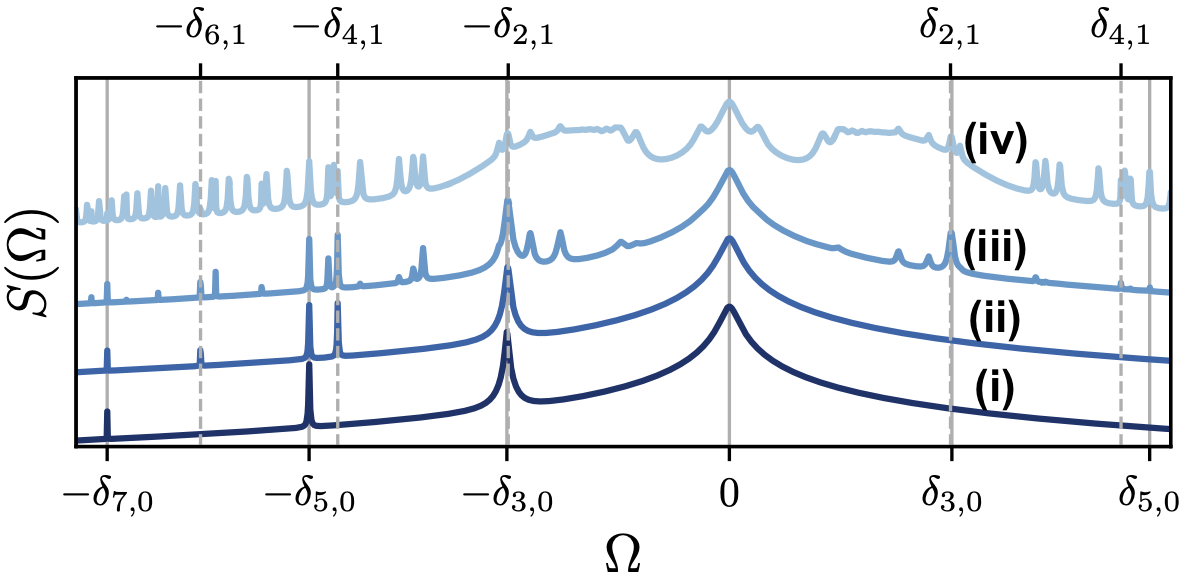}
\includegraphics[height=3.5cm]{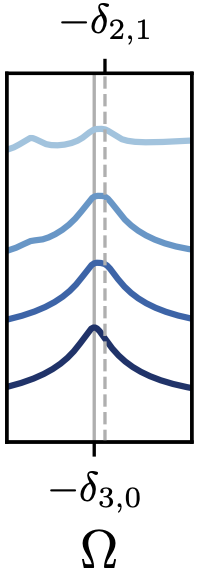}\\
\includegraphics[width=\columnwidth]{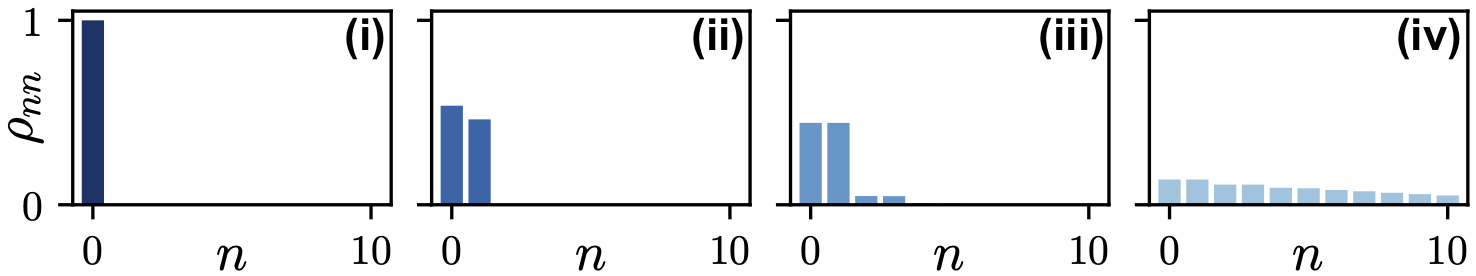}
\caption{Cavity output spectrum at different times after preparation of the cat state which correspond to different thermalization stages whose occupation pattern are shown in the lower panels with the same colors.
The curves are shifted vertically for an easier comparison.
The upper right panel gives a closer look at the first sideband.
Here, we set $\kappa = \omega$ and $g = 0.1\kappa$.
}\label{fig:spectra}
\end{figure}

In Fig.~\ref{fig:spectra} the cavity output spectrum is presented.
The plot shows that by studying the cavity output spectrum one clearly identifies a pure groundstate of the DW potential from other states.
In other words, the spectrum in Fig.~\ref{fig:spectra} shows that at the groundstate, the dark blue curve denoted as (i), one only has characteristic sideband peaks at $\Omega = -\delta_{n,0}$ with $n = 3,5,7$ and less prominently at higher odd values of $n$.
Note that the broad peak at $\Omega \approx 0$ corresponds to the lowest transition $\delta_{1,0}$ that has the largest decoherence rate $\Gamma_{1,0}$.
The central peak as well as the sidebands grow broader as the mechanical system thermalizes.
This is presented in Fig.~\ref{fig:spectra} for various thermal states of the DW potential with different effective temperatures:
Starting from the pure groundstate (i) to almost fully thermal state of (iv).
The corresponding occupation pattern of the mechanical state, $\rho_{nn}$ are presented in the lower panels.
As the system is incoherently redistributed through the thermalization process and the higher excited states get occupied extra sidebands emerge in the cavity spectrum.
At the same time the sidebands get broadened as a result of the increased thermal decoherence rate. 
%%%%%%%%%%%%%%
%%%%%%%%%%%%%%

%§§§§§§§§§§§§§§§§§§§§§§§§§§§§§§§§§§§§
\section{Summary and conclusion}
\label{sec:conclusion}
In summary, we have proposed and numerically analyzed a protocol for preparing a macroscopic spatial superposition state of a massive object.
The scheme is based on shortcut to adiabatic preparation of a system in the groundstate of a double well potential.
The counterdiabatic driving mechanism has been employed for accelerating the approach to the desired DW form and avoiding thermalization of the system.
Our results prove that very high fidelity cat states can be attained with large spatial separation by only compensating for a few diabatic transitions.
We have also proposed a setup based on the superconducting circuits with graphene for implementing the scheme.
The CD driving can be experimentally accomplished by employing driven cavity modes where each mode drives one transition, see Ref.~\cite{Abdi2016} for the details.
Given the limited number of cavity modes in a superconducting circuit our scheme can prove experimentally feasible.
The efficiency and robustness of the protocol has been benchmarked by taking into account various factors and imperfections.
The results show that the STA preparation of a macroscopic cat state of a graphene nanoresonator is robust and noise-resilient.
The mechanical state can be readout through spectroscopy of a resonantly driven cavity mode which is weakly coupled to the resonator.
As our investigations has been presented this method can efficiently fingerprint the state thanks to the high anharmonic nature of the DW potential.
Finally, it is worth mentioning that the protocol studied in this work can be combined with the dissipative approach presented in Ref.~\cite{Abdi2016} by adding one final step.
In fact, after the third step one could employ a cavity mode for sideband cooling of the $\ket{1}\hspace{-1mm}\leftrightarrow\hspace{-1mm}\ket{0}$ transition.
This transition is the most destructive thermal channel that quickly degrades the ground cat state into a mixed state of two deflections.
The cavity cooling can slow down the decoherence and give a longer coherence time which allows for better detection of the state.

\begin{acknowledgements}
The authors thank L. Pakdel and M. Fani for fruitful discussions.
This work was partially supported by the Yangyang Development Foundation.
\end{acknowledgements}

%%%%%%%%%%%%%%%%%%%%%%%
\appendix

%¶¶¶¶¶¶¶¶¶¶¶¶¶¶¶¶¶¶¶¶¶¶¶¶¶¶¶¶¶¶¶¶¶¶¶¶¶¶
\section{Mechanical properties of the graphene resonator} \label{app:elastic}
The elastic properties of a free-standing graphene membrane depends on its geometry as well as the fabrication method.
In this work we have considered a rectangular form for the membrane with dimensions $L\times w$ which has been pinned along two parallel edges.
In the case of graphene on superconducting materials, due to the large differences in the modulus of elasticity, one has a significant tensile force at those edges who hold the graphene membrane atop of the support material.
This large built-in tension imposes pinned boundary conditions on the mechanical resonator.
In other words, the same as those for a string fixed at its both ends.
Therefore, the mode profiles are $\varphi_n(x)=\sin(n\pi x/L)$ and the frequencies $\omega_n=\sqrt{\mathcal{T}/\mu}(n\pi/L)$ with $n=1,2,\cdots$.
That is, the membrane deflection can be cast in the form of $z(x,t)=\sum_n u_n(t)\varphi_n(x)$ with the mode amplitudes $u_n(t)$.
Here, $\mathcal{T}$ and $\mu$ are the tensile force at the boundaries and the two-dimensional mass density of the membrane, respectively.

The bending modes cause a small extension in the length of the resonator and consequently give rise to a nonlinearity in the system.
This extra tension is $\Delta \mathcal{T}=Yh(\Delta L/L)$, where $Y$ is the Young modulus and $h$ is the thickness of the membrane.
The total length stretch is given by $\Delta L=\frac{1}{2}\int_0^Ldy|\partial_y z(y,t)|^2$.
This brings us at $m_n=\frac{1}{2}\mu Lw$ for the mass and $\beta_n = (Yhw/8L^3)(n\pi)^4$ as the Duffing nonlinearity of the $n$th mode with $w$ the width of membrane.
These relations has been used in the text for computing the system parameters.

%¶¶¶¶¶¶¶¶¶¶¶¶¶¶¶¶¶¶¶¶¶¶¶¶¶¶¶¶¶¶¶¶¶¶¶¶¶¶
\section{Electrostatic potential} \label{app:electrostatic}
To construct an anti-parabola that mimics symmetry of the membrane fundamental mode we propose to employ two line electrodes at $ x = \pm b $ with length $2a$ and symmetrically positioned beneath the center of the membrane, see Fig.~\ref{fig:setup_DW}.
Such configuration can roughly by estimated by two infinitesimally thin rods.
Hence, the resulting electrostatic potential at the point $ (x,y=0,z) $ is given by
\begin{align*}
V_{\rm e}(x,z) =& V_0 \Big(\ln [\frac{\sqrt{a^2 +(x-b)^2 + (z-z_0)^2} +a}{\sqrt{a^2 +(x-b)^2 + (z-z_0)^2} -a}] \\
&~~+ \ln [\frac{\sqrt{a^2 +(x+b)^2 + (z-z_0)^2} +a}{\sqrt{a^2 +(x+b)^2 + (z-z_0)^2} -a}] \Big) ,
\end{align*}
where $V_0$ is the potential applied to the electrodes and $z_0$ is the equilibrium distance of the membrane from the surface that includes the electrodes.

\begin{figure}[tb]
   \includegraphics[width=0.9\columnwidth]{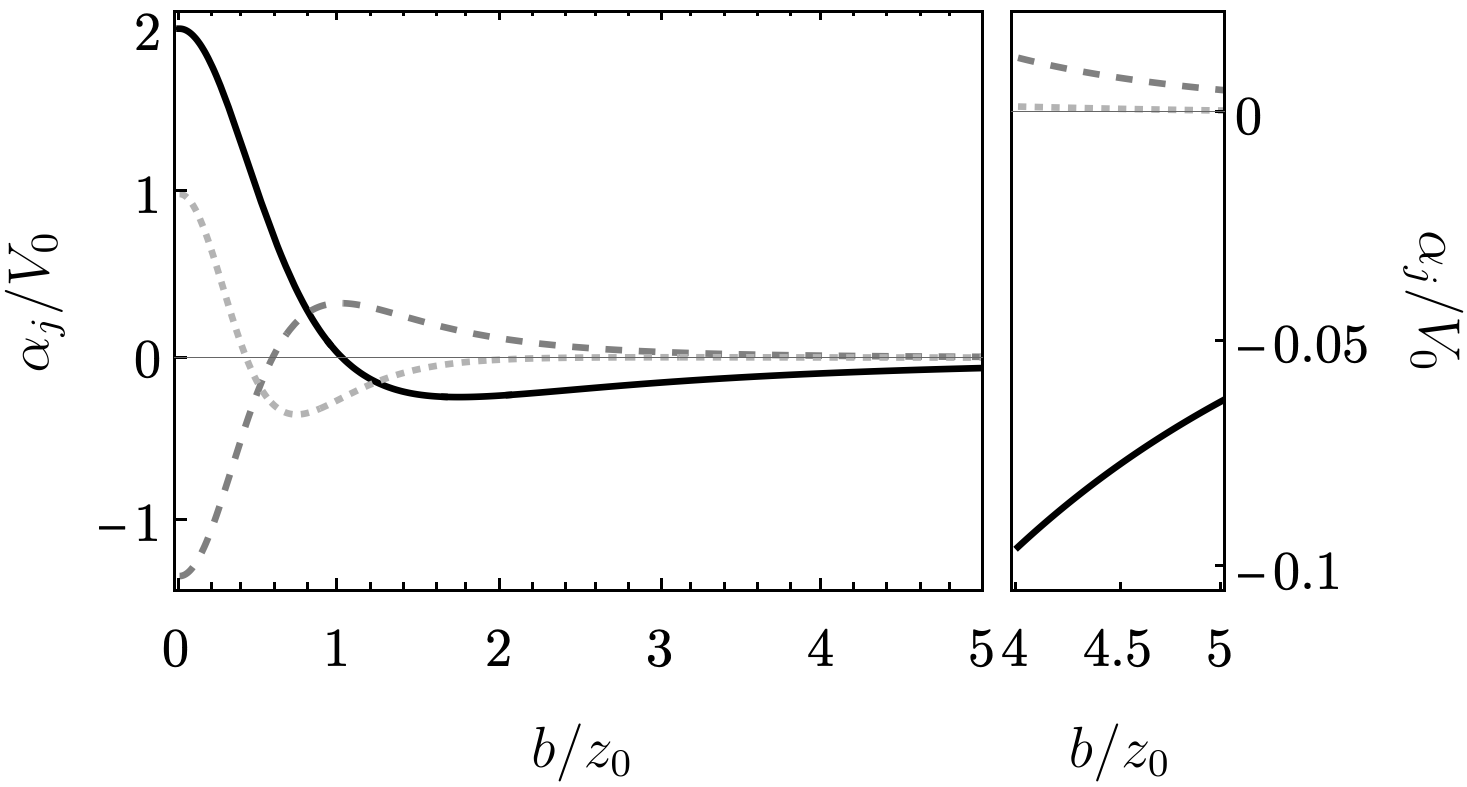}
   \caption{Coefficients of expansion of the electrostatic potential $\alpha_j$ as a function of their separation: $j=2$ (bold line), $j=3$ (dashed line), and $j=4$ (dotted line).
   Here, we have set $a=10 z_0$.
   The right panel gives a closer look at the large $b/z_0$ values.
   Note that $\alpha_4 \approx 0$ at this regime.}
\label{fig:Ves}
\end{figure}

By assuming small vibrational amplitudes one Taylor expands $V_{\rm e}$ around $ z = 0 $ at $ x=y = 0 $ to find the effective external potential `felt' by the membrane: $V_{\rm e}(z) = \sum_j \alpha_j z^j$.
The linear contribution of this external potential leads to a shift in the equilibrium position of the membrane.
This, in turn, can be compensated for by the gate potential if necessary.
However, the higher expansion terms contribute to the total Hamiltonian of the system and determine its dynamics.
In Fig.~\ref{fig:Ves} the coefficients of expansion $\alpha_j$ are plotted against $b$ for $a=10z_0$.
At $b = z_0 /\sqrt{3}$ one has $\alpha_3 = 0$ which, in principle, is desirable since the potential remains symmetric under parity transformation, when neglecting the higher order odd terms.
%%%%%%%%%%%%%%
%%%%%%%%%%%%%%

Nonetheless, the membrane capacitively couples to the superconducting circuit and a considerable electromechanical coupling rate, which is necessary for the initial sideband cooling, demands a considerable capacitance.
That is, a small spacing between the membrane and the gate electrode and yet a large area overlap between them (see the red rectangular electrode in Fig.\ref{fig:setup_DW}).
Therefore, $b=z_0/\sqrt{3}$ does not fulfill our requirements for the considerable electromechanical coupling.
Instead, we consider the case of $ b \gg z_0 $ were $\alpha_2$ remains the dominant term.
The negligibility of the higher order terms is justified because of the quantum regime that we are interested in.
In fact, in the Hamiltonian one has $\hat{z} = z_{\rm zpm}(\hat{b} +\hatd{b})$, where $z_{\rm zpm} = \sqrt{\hbar/2m\omega}$ is the zero point amplitude.
For the parameters discussed in our setup this is $z_{\rm zpm} \sim 1$~pm.
Hence, the contribution from the higher order expansion terms are further suppressed in the quantum regime.

%¶¶¶¶¶¶¶¶¶¶¶¶¶¶¶¶¶¶¶¶¶¶¶¶¶¶¶¶¶¶¶¶¶¶¶¶¶¶
%\section{Implementation of the counterdiabatic drives} \label{app:implementation}
%The counterdiabatic drives in our proposed setup can be implemented through the appropriate manipulation of the superconducting cavity modes.
%Here, we only consider one cavity mode for the sake of simplicity.
%The formulation can be generalized to multiple cavity modes.
%In our proposed setup, the cavity modes capacitively couple to the vibrational degree of freedom of the graphene membrane.
%The effective electromechanical Hamiltonian at every instance of time during the ramp is given by Eq.~\eqref{Hopto}.
%By expressing the position operator in terms of the $\hat{H}_{\rm DW}$ eigenstates one gets
%\begin{equation}
%	\hat{H}_{\rm tot} = -\hbar\Delta \hatd{a}\hat{a} +\hat{H}_{\rm DW} +\hbar g \sum_{m,n} z_{mn}\tran{m}{n}(\hat{a} +\hatd{a}).
%\end{equation}
%In the interaction picture of $\hat{H}_0 = -\hbar\Delta \hatd{a}\hat{a} +\hat{H}_{\rm DW}$ the Hamiltonian reads
%\begin{equation}
%	\tilde{H}_{\rm tot}(t) = \hbar g\sum_{m,n}z_{mn}(e^{-i(\delta_{mn} -\Delta)t}\hat{a} +e^{-i(\delta_{mn} +\Delta)t}\hatd{a})
%\end{equation}

%%%%%%%%%%%%%%%%%%
\begin{figure}[tb]
\includegraphics[width=0.8\columnwidth]{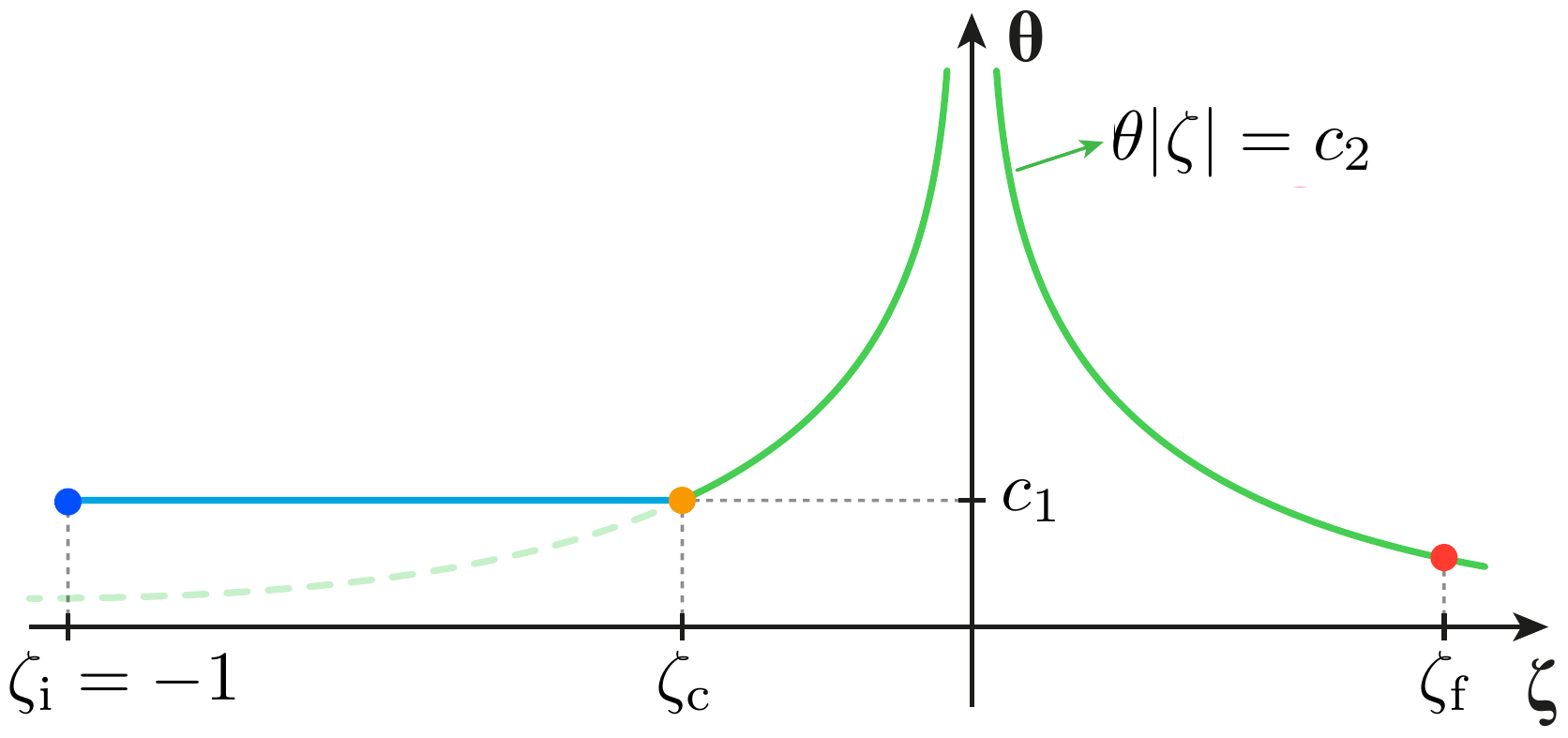}
\includegraphics[width=\columnwidth]{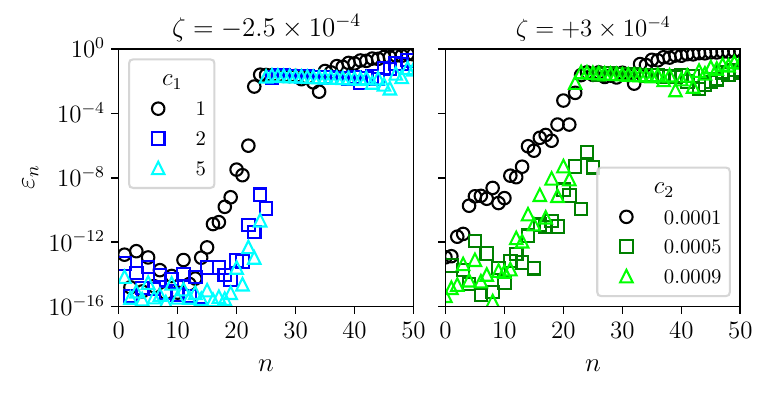}
\caption{The top panel: $\theta$ as a function of $\zeta$.
In the lower panels the relative error for the eigenvalues of the truncated Hamiltonian are given for different values of $c_1$ for $\zeta=-2.5\times 10^{-4}$ (left) and different values of $c_2$ for $\zeta=+3\times 10^{-4}$ (right).}
\label{fig:error}
\end{figure}
%%%%%%%%%%%%%%%%%%

%¶¶¶¶¶¶¶¶¶¶¶¶¶¶¶¶¶¶¶¶¶¶¶¶¶¶¶¶¶¶¶¶¶¶¶¶¶¶
\section{Numerical method}\label{app:nuermical}
The continuous variable nature of the system gives a infinite Hilbert space whose numerical analysis demands for truncations.
To avoid the burden of expensive computations due to considering a large Hilbert space, one needs to choose the computational basis carefully.
It is easy to check that eigenbasis of the `original' harmonic oscillator with frequency $\omega$ only gives reliable results for very high Hilbert space truncations.
By investigation, we find that a harmonic oscillator basis with $\omega_0 = \sqrt{\theta \nu/m} = \omega\sqrt{\theta \left| \zeta \right| }$ (for $\zeta \geq -1$ and $\theta > 0$) gives much better results for a proper choice of $\theta$.
Hence, we write
\begin{subequations}
\begin{align}
\hat{z} &= \sqrt{\frac{\hbar}{2m\omega_0}}(\hat{b}+\hat{b}^\dagger), \\
\hat{p} &= -i \sqrt{\frac{m\hbar\omega_0}{2}}(\hat{b}-\hat{b}^\dagger),
\end{align}
\end{subequations}
where $\hat{b}$ ($\hatd{b}$) is the bosonic annihilation (creation) operator with the commutator $[\hat{b},\hatd{b}]=1$.
By plugging into \eqref{eqn:mechamil} we arrive at the Hamiltonian
\begin{equation}
\frac{\hat{H}_{\rm DW}}{\hbar\omega_0} = -\frac{(\hat{b}-\hat{b}^\dagger)^2}{4}  - \frac{\text{sgn}(\zeta) (\hat{b}+\hat{b}^\dagger)^2}{4 \theta} + \frac{\gamma (\hat{b}+\hat{b}^\dagger)^4}{(\theta |\zeta|)^{3/2}},
\label{eqn:HDW_harmonic_basis}
\end{equation}
where $\gamma = \beta \hbar / (16 m^2 \omega^3)$ has been introduced.
%We also omitted to write the time-dependence of $\hat{\mathcal{H}}_{0}$ and $\zeta$.

Note that since in Eq.~\eqref{eqn:HDW_harmonic_basis} $\zeta$ appears in the denominator of the last term one has to be meticulous when dealing with the small values of $\zeta$.
In our numerical analysis, we find by inspection that for the range of $\zeta \in [-1, -2.5\times 10^{-4}]$ a constant value of $\theta = c_1$ gives results with high precision for a Hilbert space truncated at $\text{dim}=50$ when $c_1$ is carefully determined.
For the remaining range including the final value $\zeta_{\rm f}=+3\times 10^{-4}$, we instead tune $\theta$ such that the denominator remains finite.
In other words, we set $\theta |\zeta| = c_2$, where again the optimal value of $c_2$ is found numerically.

Now we discuss the method we used to find the optimal values of $c_1$ and $c_2$.
The goal is to have smallest truncated Hilbert space, yet with high precision.
We do this by contrasting the eigenvalues obtained from a basis with two truncated dimensions: one high ($\text{dim}=1000$) and the other low ($\text{dim}=50$) for different values of $c_1$ and then $c_2$.
We indicate the former by $E_n^{\rm H}$, while the latter is indicated by $E_n^{\rm L}$.
A higher truncation of the Hilbert space always gives more reliable results for the states with lowest eigenvalues.
Therefore, they give a good reference for gauging the accuracy of eigenstates in lower Hilbert space truncations.
Hence, we define the relative error as $\varepsilon_n = |E_n^{\rm H} - E_n^{\rm L}|/|E_n^{\rm H} + E_n^{\rm L}|$, with $n=0,1,2,...$ being indexing the energy levels.

For the first part that contains large values of $|\zeta|$, i.e. $\zeta \in [-1, -2.5\times 10^{-4}]$, by trying different values for $c_1$ and comparing the errors, we find that $c_1 = 2$ gives an energy spectrum with high accuracy for up to the 25th level.
%Note that, this is indeed the best result one can expect for a truncated Hilbert space.
%That is, when Hamiltonian of a continuous variable system is truncated at $\text{dim}$, only eigenvalues and eigenstates with $n<\text{dim}/2$ can be correct.
For the second part we find $c_2 = 5\times 10^{-4}$ giving the least error for states with $n \leq 25$, see Fig.~\ref{fig:error}.

Having a reliable method for finding the eigenstates and eigenvalues of the double-well Hamiltonian for all values of $\zeta$ in the interval of interest, we now turn to the numerical method we have used for studying the dynamics of our system.
The full system dynamics is given by the quantum optical master equation \eqref{eqn:master}, where the Hamiltonians as well as the collapse operators in the dissipation are a function of time through the time-dependence of the control parameter $\zeta(t)$.
Having the eigenstates and eigenvalues of $\hat{H}_{\rm DW}(\zeta)$ through the numerical computations, we generate the counterdiabatic drive Hamiltonian $\hat{H}_{\rm drv}(\zeta)$ by plugging the parameters and the eigenstates in Eq.~\eqref{eqn:HCD_even}.
Similarly, we generate the collapse operator $\hat{A}(\zeta)$ function.
Eventually, we employ the QuTiP package in python for finding the numerical solution of Eq.~\eqref{eqn:master}~\cite{Johansson2013}.

\bibliography{stadw}

\end{document}